\documentclass[final,twocolumn,sort&compress]{elsarticle}
\usepackage[utf8]{inputenc}
\usepackage[tbtags]{amsmath}
\usepackage{amssymb,amsthm,amsfonts,relsize,layout,indentfirst,graphicx,float,wrapfig,color,tabls}
\usepackage{boxedminipage,array,lineno}
\usepackage[font={bf,small},textfont={it,md,small}]{subfig}
\linenumbers
\begin{document}
%
%
	\title{Bayesian inference of inaccuracies in radiation transport physics from inertial confinement fusion experiments}
	\tnotetext[t1]{This work performed under the auspices of the U.S. Department of Energy by Lawrence Livermore National Laboratory under Contract DE-AC52-07NA27344. LLNL-JRNL-617033}
	\author[llnl]{J.A.~Gaffney\corref{cor1}}
	\ead{gaffney3@llnl.gov}
	\author[llnl]{D.~Clark}				
	\author[llnl]{V.~Sonnad}
	\author[llnl]{S.B.~Libby}
	\cortext[cor1]{Corresponding author}
	\address[llnl]{Lawrence Livermore National Laboratory, 7000 East Ave, Livermore, CA 94550}
%
%
%
%
	\begin{abstract}
	First principles microphysics models are essential to the design and analysis of high energy density physics experiments. Using experimental data to investigate the underlying physics is also essential, particularly when simulations and experiments are not consistent with each other. This is a difficult task, due to the large number of physical models that play a role, and due to the complex (and as a result, noisy) nature of the experiments. This results in a large number of parameters that make any inference a daunting task; it is also very important to consistently treat both experimental and prior understanding of the problem. In this paper we present a Bayesian method that includes both these effects, and allows the inference of a set of modifiers which have been constructed to give information about microphysics models from experimental data. We pay particular attention to radiation transport models. The inference takes into account a large set of experimental parameters and an estimate of the prior knowledge through a modified $\chi^{2}$ function, which is minimised using an efficient genetic algorithm. Both factors play an essential role in our analysis. We find that although there is evidence of inaccuracies in off-line calculations of X ray drive intensity and Ge $L$ shell absorption, modifications to radiation transport are unable to reconcile differences between 1D HYDRA simulations and the experiment.
	\end{abstract}
%
%
	\maketitle
%
%
\section{Introduction}
In recent inertial confinement fusion (ICF) experiments performed at the national ignition facility (NIF) \cite{nif}, significant differences between radiation-hydrodynamic simulations and experimental data have been observed \cite{landen12}. It is not clear whether these simulations are inaccurate, or that they neglect some important physical effect. It is quite challenging to investigate which aspects of physics models are causing discrepancies and should be improved, largely due to the complex, nonlinear dependance of ICF capsule evolution on a large number of underlying models.

%
The complex nature of the experimental designs is an important source of the difficulties. There are a large number of experimental parameters that are only known with limited accuracy; variations in these parameters represent a noise source in the experimental data that can reduce the significance of the experimental result. Since the physical models we aim to investigate are fairly well constrained by a large amount of previous work, it is important to take into account the relative significance of the experiment and of any previous work. The interplay between experimental and previous information is an essential ingredient in a reliable analysis, and is often neglected. Its inclusion requires a consistent treatment of all physical and experimental parameters; together there are far too many of these to treat directly, however they are too important to neglect completely.\par
%
%
In this paper we present an analysis of experimental data taken from a single NIF shot, N110625. The aim is develop a method of investigating microphysics models taking into account many of the noise sources in the experiment, and prior work. We use an inference model that has been developed specifically to allow the large number of parameters to be dealt with in a consistent manner \cite{gaffney12_bayesian_submitted}. In this work we focus on inferring information about radiation transport in the ablator of an ICF capsule from time resolved data taken from radiography \cite{hicks12}. Radiation transport relies on several physics models which must be approximated to make a full capsule simulation tractable, and as a result are often considered to be potential sources of model inaccuracy. In this work existing microphysics tables are modified in physically motivated ways; these modifiers are interpreted as measures of the inaccuracies in the physics models, and their inferred values give information about the source of difficulties in describing experimental observations.\par
%
%
\section{Microphysics in Inertial Confinement Fusion ablators}
In a typical indirect drive ICF design a spherical plastic shell, filled with deuterium-tritium (DT) fuel, is bathed in X rays created by the interaction of laser light with a high $Z$ hohlraum. The resulting ablation of the outer plastic produces a rocket action that implodes the shell, compressing the fuel until it undergoes thermonuclear fusion. The propagation and absorption of X ray energy in the plastic and fuel is an essential piece of describing the implosion that requires detailed models of microscopic physics. These physics issues are described by a suit of computer simulations which provide, for example, tables of radiative opacities which are taken as input by subsequent radiation-hydrodynamic simulations \cite{castor}.\par
Achieving ignition is a challenge and so the design of successful targets requires careful tuning of a large set of design parameters, based on the results of simulations \cite{haan11}. This means that the fine details of microphysics models can be very significant; nevertheless, the important aspects can be understood with relatively simple one dimensional models \cite{atzeni}. We will discuss the important aspects of microphysics models in these terms.\par
At its peak, the X ray drive on the outer surface of the capsule has a brightness temperature of around 300eV. The majority of the energy of this field is in photon energies that coincide with the $K$ shell absorption edge in carbon (which accounts for $\sim$50\% by number of the plastic ablator) and so the model of this absorption feature plays an important role in determining energy deposition in the ablator. Higher photon energies are able to propagate all the way through the carbon, depositing their energy in the DT fuel. Heating of the fuel by these hard X rays has a detrimental effect on the implosion since, for the efficient adiabatic implosions driven by the NIF, the final density is in part determined by the initial temperature of the DT. Preheat by X rays reduces fuel compressibility and ultimately reduces the final convergence that can be acheived. An important player in this preheat is emission from the $M$ shell of the gold hohlraum wall, which produces an enhancement over the thermal specturm of photon energies $>1.8$KeV; in order to block these from reaching the fuel a dopant layer is buried in the ablator. In this work we consider germanium doped ablators, in which case absorption by the Ge $L$ shell aligns with the Au $M$ shell emission and prevents preheat of the fuel. In reality, the growth of 2 and 3 dimensional instabilities also plays a very important role in determining the implosion efficiency. In severe cases these can be much more important than the 1D considerations that we have described.\par
These two aspects of radiation transport, namely absorption of the drive field and preheat of the fuel, clearly depend on models of the generation of the drive spectrum and of the absorption in carbon and germanium at a large range of conditions (10-200 eV, 1-10 g/cc). They also have direct consequences for the dynamics of the implosion. A simple rocket model for the inwards acceleration of the ablator \cite{atzeni} shows that the velocity and remaining ablator mass are directly related to velocity of the ablated material, and therefore the absorption of drive radiation. The density of the fuel at a given time is related to the preheat. Measurements of these three quantities, as described in section \ref{sec:results}, can therefore provide information about underlying radiation transport physics models. The complexity of ICF experiments and radiation-hydrodynamic simulations means that extracting this information is a challenging data analysis problem; we describe a method of performing this analysis in the next section.\par
%
%
\section{Bayesian analysis of ICF experiments}
The relationship between physical models, which themselves are very complex, and the data is approximated by radiation-hydrodynamic simulations which may not be well behaved enough to allow the use of computational inversion techniques \cite{hanson98} or fitting techniques \cite{sacks89,kennedy01}; the large number of physical models that control the evolution of an ICF target also presents a problem for these methods. The complex nature of the experiments also means that there are a large number of experimental parameters; although these are often constrained by target metrology and design tolerances, their large number makes them a significant source of noise in simulations \cite{clark-pres11}. Dealing with the very large space of physical and experimental parameters is an important challenge to a consistent analysis of ICF data. The usual methods of reducing the number of parameters, for example by Monte-Carlo sampling (see, for example, \cite{roe07}), are prohibitively expensive, and simply neglecting parameters will lead to misleading results.\par
In \cite{gaffney12_bayesian_submitted} we have developed an inference method that allows these problems to be addressed. The approach is to separate out those parameters that are known to affect radiation-hydrodynamic simulations but are not of direct interest to the investigation of microphysics models; these are defined as `nuisance parameters'. Typically these parameters refer to experimental variables which have a known probability distribution, for example a target dimension that has been measured with some error bar. The probability distributions of all nuisance parameters are mapped onto the output of radiation-hydrodynamic simulations; as a result the simulation output can be considered as being probabilistic. In our model we assume a linear response to nuisance parameters, resulting in an analytic expression for the probability distribution of simulation outputs (the \emph{likelihood}). Parameters that are physically interesting (and therefore will be inferred from experimental data), such as microphysics models, are kept separate from the nuisance parameters allowing their relationship with experimental data to be described using the full complexity of the simulation code.\par
The inference model we have outlined is based on the maximum \emph{a posteriori} (MAP) estimate; that is, the most probable values of all parameters of interest when the experimental data and prior have been taken into account. In our analysis these values are found by minimising the function \cite{gaffney12_bayesian_submitted}
\begin{equation}
	\begin{split}
		I(\theta|d_{exp})=& \sum_{i} \frac{(d_{exp,i}-d_{m}(\theta)_{i})^{2}}{\sigma_{exp,i}^{2}}  \\
			 & - (d_{exp}-d_{m}(\theta))^{T} \beta^{T}\beta (d_{exp}-d_{m}(\theta)) \\
			 & +\frac{1}{2}{\rm ln}\left(|\Lambda_{\eta}||\alpha^{T}\alpha|\right) - {\rm ln}P(\theta)
	\end{split}
	~{\rm }
	\label{eq:information_likelihood_marginal}
\end{equation}
with respect to the vector of interesting parameters $\theta$. In the above expression, $d_{m}(\theta)$ is the vector of simulation outputs for given interesting parameters and nominal values of the nuisance parameters, $d_{exp}$ is the vector of experimental data, $P(\theta)$ is the prior distribution of interesting parameters (discussed below) and the matrices $\alpha$ and $\beta$ satisfy the equations
\begin{align}
	\alpha^{T}\alpha &=A^{T}\Lambda_{exp}^{-1}A+\Lambda_{\eta}^{-1} \notag \\
	\beta^{T}\alpha &= \Lambda_{exp}^{-1}A \notag
	~{\rm .}
\end{align}
These matrices summarise the effect of nuisance parameters on our analysis; $\Lambda_{exp}$ and $\Lambda_{\eta}$ are the covariance matrices of the experimental measurement and nuisance parameters, respectively, and $A$ is the linear response of the simulation to nuisance parameters $\eta$: $A_{ij} = \frac{\partial d_{m}(\theta)_{i}}{\partial \eta_{j}}$.\par
Equation \eqref{eq:information_likelihood_marginal} takes the form of a modified $\chi^{2}$ function. The first term on the right hand side is the usual $\chi^{2}$ analysis, and the second can be interpreted as a loss of information from the experiment due to nuisance parameters. The third is a normalisation factor. The final term expresses the contribution from prior work on the values of the interesting parameters. In our application we interpret this term as an estimated error bar on the physical models we aim to investigate, reflecting previous work to validate them. The inclusion of this prior information provides context for the experimental result, allowing inferences to be obtained from a single observation. In \cite{gaffney12_bayesian_submitted} this was shown to play a very important role in the analysis of NIF data.\par
The summary of nuisance parameters in the matrix $\beta^{T}\beta$ has reduced the number of variables we must consider to only the ones of direct interest. The resulting smoothing of the simulation output also means that the minimisation of equation \eqref{eq:information_likelihood_marginal} can be approached using standard numerical methods. In this work we use a genetic algorithm (GA) to efficiently perform the minimisation. The details of the genetic algorithm have been optimised to allow an efficient exploration of a large parameter space; the sacrifice is that the algorithm is more likely to find local minima. In the case of the ICF data we will consider here, this is not expected to be an issue since the interplay between likelihhod and prior tends to produce a single minimum. In more complex cases this can be tested by using several random initialisations, or avoided by using a more robust algorithm.\par
%
%
\section{Application to NIF experimental data} \label{sec:results}
We aim to demonstrate the application of our Bayesian inference method to the investigation of microphysics models using NIF data. We use 1D simulations of a capsule implosion performed using the HYDRA radiation-hydrodynamics code \cite{hydra}. Our investigation proceeds by defining a set of modifiers to the inputs of these simulations, and inferring the values of these modifiers. We are concentrating on physics issues in radiation transport and so our modifiers are to the X ray drive spectrum impinging on the capsule's outer surface (found from seperate models of the hohlraum), and to relevant opacity models of the ablator material (taken from the TABOP opacity model). The use of modifiers, placed on the results of existing calculations, allows our inference results to be interpreted as implied inaccuracies in microphysics models. We give details of our modifiers in table \ref{tab:modifiers}. In the case of the drive timing modifier, the prior error bar reflects the error bar on the DANTE instrument \cite{kline10}, which gives a time-resolved measurement of the drive radiation temperature. This instrument has played an important role in the development of the separate hohlraum simulations which produce drive profiles for our capsule simulations. For all other modifiers, prior errors are estimated in order to reflect the expected accuracy of the underlying physical models. All modifiers, with the exception of the drive timing, are dimensionless multipliers on existing models and so their `nominal' (and therefore prior) values are 1; the drive timing has a nominal shift of 0 ns.\par
\begin{table*}
\centering
\begin{tabular}{|m{2.5cm}|m{5.5cm}|m{6.0cm}|m{2cm}|}
 	\hline
	\centering{Modifier Name} & \centering{Description} & \centering{Expected Effect} & Prior Error \\
	\hline
	Drive Intensity & Multiplies intensity of $4^{\rm th}$ rise in X ray drive 
			& Increased drive results in increased velocity and decreased ablator remaining at given time & $\pm0.1$ \\ \hline
	Drive Timing	& Shifts the timing of the $4^{\rm th}$ rise of the X ray drive 
			& Earlier rise increases drive at given time & $ \pm0.1$ \\ \hline
	Au $M$ Shell	& Multiplies the intensity of the gold $M$ shell component of X ray drive spectrum 
			& Increased $M$ shell results in increased preheat and reduced $\rho R$ at given time & $\pm 0.2$ \\ \hline
	C $K$ Edge 	& Multiplies the opacity of the $K$ shell absorption edge in carbon 
			& Increased absorption increases effective drive & $\pm 0.1$ \\ \hline
	Ge $L$ Edge	& Multiplies the opacity of the $L$ shell absorption edge in germanium 
			& Increased absorption reduces preheat & $\pm 0.1$ \\
	\hline
\end{tabular}
\caption{Description of the modifiers placed on input physics models. The values of these modifiers are inferred from experimental data using the method described in the text, and are intended to give information about the accuracy of radiation transport models for NIF ablators.}
\label{tab:modifiers}
\end{table*}
Experimental data are taken from a single NIF `convergent ablator' shot, N110625. This experiment utilised a germanium doped capsule which was radiographed as it imploded giving a time- and space- resolved measurement of plasma density \cite{hicks10,hicks12}. This then gives time-resolved data for the implosion velocity, mass of the ablator, and the $\rho R$ product of the imploding fuel shell. We consider these three data points, taken at three times during rocket-like phase of the implosion, in our analysis. The use of implosion velocity and ablator mass, which diagnose the drive, along with the $\rho R$ which is sensitive to preheat of the fuel, should allow the degeneracy of our modifier set (for example the drive intensity and C $K$ shell) to be lifted. This is important since such degeneracy results in a set of multiplier values that minimise equation \eqref{eq:information_likelihood_marginal}; the inclusion of the $\rho R$ data should select a single one of these values since it more fully reflects the physics of the problem.\par
For the multipliers and experimental data described, our genetic algorithm is randomly initialised and procedes by automatically calling HYDRA. The nuisance parameter modification $\beta^{T}\beta$ is calculated for the 29 physical dimensions, densities and material composition parameters of the capsule \cite{haan11}, which are assumed to be known with an error of 1\%. We ran the GA for 25 generations with 92 members per generation, requiring up to 2300 HYDRA simulations (the actual number is lower due to the optimisations made to the GA), equivalent to $<200$ CPU hours. In table \ref{tab:results} we give the position of the results for two cases; including and neglecting the prior, respectively. Since the position of the minimum of equation \eqref{eq:information_likelihood_marginal} is determined by the relative importance of the prior and experimental results, comparison of these two cases provides information about the significance of the experiment in measuring radiation transport physics.\par
\begin{table}
\centering
\begin{tabular}{|c|c|c|}
	\hline
	Modifier	& No Prior	& Including Prior	\\
	\hline\hline
	Drive intensity	& 0.57		& 0.90		\\
	Drive timing	& -0.45 ns	& 0.01 ns	\\
	Au $M$ shell	& 1.84		& 0.97		\\
	C $K$ edge	& 0.92		& 1.0		\\
	Ge $L$ shell	& 1.15		& 1.16		\\
	\hline
\end{tabular}
\caption{Positions of the best fit to experimental data for NIF shot N110625. In both cases 29 nuisance parameters are included; comparison of the two sets of data measures the significance of the experimental data when compared to prior knowledge.}
\label{tab:results}
\end{table}
In table \ref{tab:results} the fit given in the `No Prior' column corresponds to a maximum likelihood (ML) analysis, in which the experimental data are the only source of information about the values of the modifiers. In this case, the results demonstrate that in order to fit the data all modifiers should be significantly different from their nominal values; this implies that microphysics models are in considerable error. Given the extensive work that has been undertaken on these models in the past, it is unlikely that this is truly the case. The previous work is taken into account in the `Including Prior' column, and the large difference in results demonstrates the importance of including prior knowledge. In that case (corresponding to the MAP result) all modifiers are much closer to their nominal values. The noise in the experiment makes the observed data insensitive to the details of radiation transport; only the overall drive and Ge absorption are significantly modified from their prior values. Our results suggest that the calculated drive intensity is too high, consistent with previous work on ICF data, and that the calculated absorption by the germanium dopant layer is too low.\par
\begin{figure}
	\centering
 	\subfloat[Implosion velocity]{\includegraphics[scale=0.25]{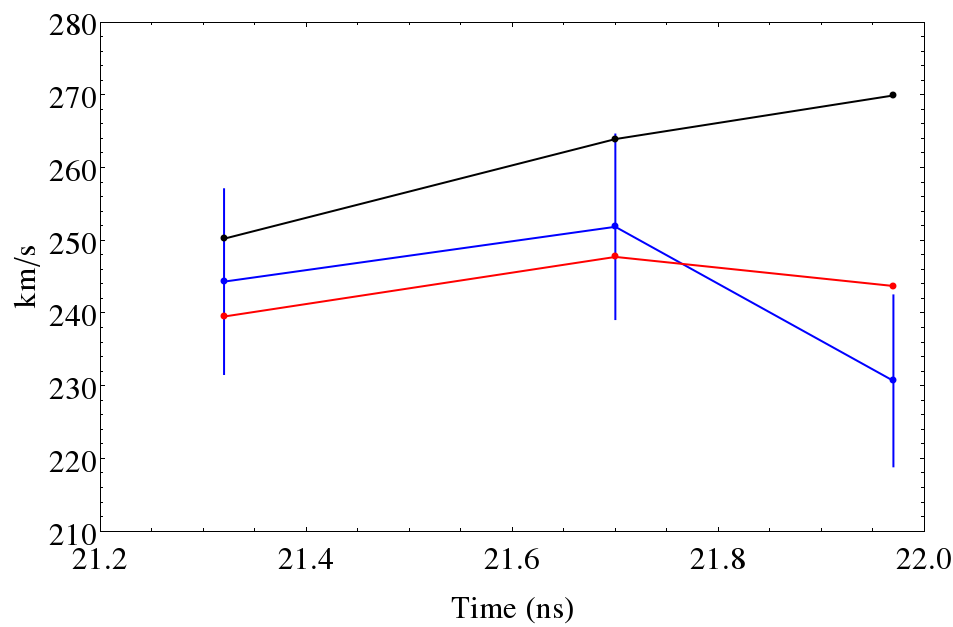}}\\
	\subfloat[Fraction of ablator remaining]{\includegraphics[scale=0.25]{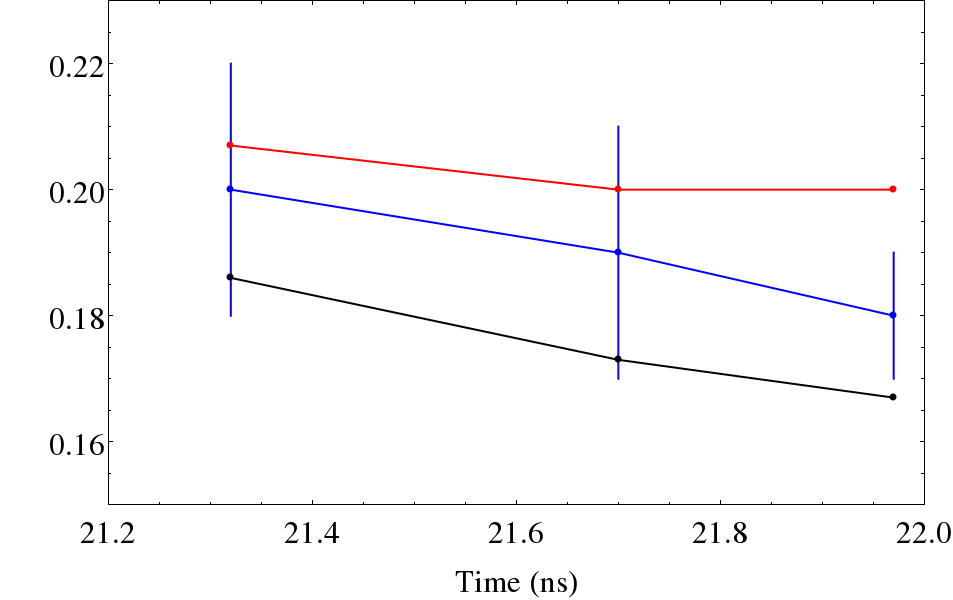}}\\
 	\subfloat[Fuel $\rho R$]{\includegraphics[scale=0.25]{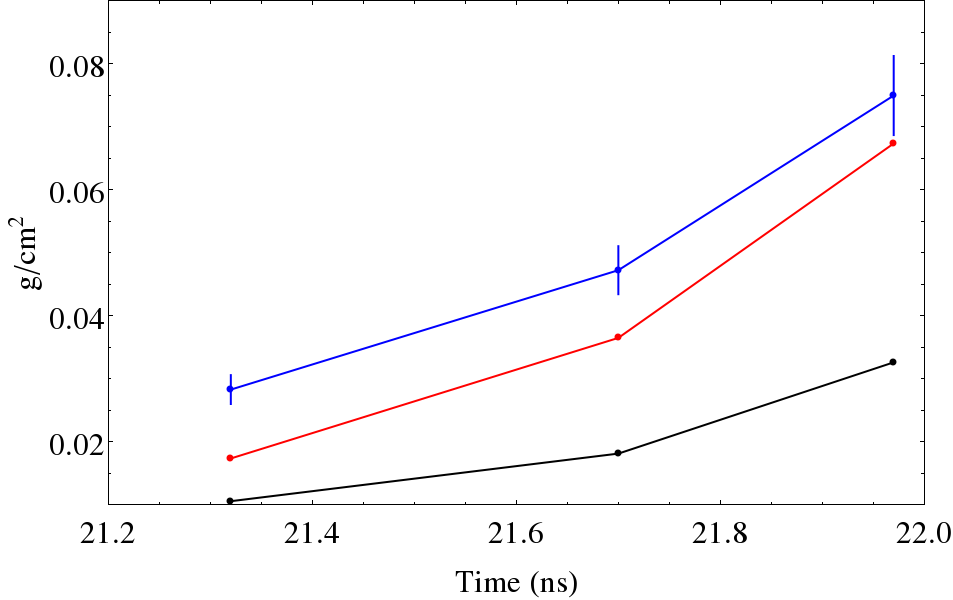}}\\
	\caption{Best fits to experimental data, corresponding to HYDRA simulations using modifiers given in table \ref{tab:results}. Experimental data are shown in blue, and inference results neglecting and including prior knowledge are shown in red and black, respectively.}
	\label{fig:results}
\end{figure}

Comparison of the best fits to experiment, found using the two inference methods (neglecting and including the prior), allows us to measure the ability of inaccuracies in radiation transport to explain problems with modelling of the experiment. The quality of the inferred fits to experimentally inferred implosion velocity, ablator fraction, and line density are shown in figure \ref{fig:results}(a),(b) and (c) respectively. In these figures, experimental data as a function of time are shown in blue, and simulation results using modifier values from table \ref{tab:results} are plotted in red (no prior) and black (prior included). The ML analysis, neglecting the prior, gives a reasonable qualitative fit to the data, but does not match within all error bars. The MAP result is much closer to an unmodified simulation and gives a poorer agreement with experiment. The inability of either approach to give a good match to the data suggests that discrepancies between simulations and experiments are not solely due to issues with radiation transport.\par
%
%
\section{Discussion and Conclusions}
The work presented in this paper demonstrates a method for inferring information about first principles physics models from ICF data. The inference model we use allows the inclusion of a large number of nuisance parameters; these play an important role in determining the information in the experimental result. This is essential when comparing experimental results with the results of previous work, which is often the case in high energy density physics. Although we focus here on radiation transport in ICF ablators, the issues we discuss are important in many of the experiments performed in high energy density physics, and the inference method we describe is easily applicable to any of these.\par
The main result of this work is that prior knowledge about microphysics plays a very important role. Including this in a consistent manner allows meaningful information to be extracted from data, so that when data imply a modification to physics models the result truly reflects the state of the art. We have also shown that the complex nature of ICF experiments means that the neglect of nuisance parameters and/or prior information in a simple $\chi^{2}$ or maximum likelihood analysis will give misleading results. In this work 29 parameters have been varied by $1\%$ in order to produce the information loss due to nuisance parameters; for the well characterised targets used at the NIF certain capsule dimensions are known to a much better level than this, however prior knowledge will play an essential role regardless.\par
Once these factors are accounted for, there is evidence that both the overall X ray drive and the absorption of the germanium $L$ shell are inaccurate. This could serve to focus subsequent investigation of the underlying models (for example further inferences of inaccuracies in charge state balance), however the poor agreement between the current best fit and the experimental data shows that issues with radiation transport cannot explain discrepancies between the details of ICF implosions and simulations. It is important to note that inferences based on an incomplete set of modifiers, which appears to be the case here, may never give a good fit to data. Until a good fit is found the physical meaning of multipliers is limited, and inferred values should be treated accordingly.\par
The method used here has been specifically designed so that an analysis with a large enough set of modifiers is feasible. Cases with 1-2 orders of magnitude more evaluations of $\chi^{2}$ are possible with a fairly modest computational requirement, and the number of nuisance parameters can be increased in our linear model with almost no numerical overhead.\par
Genetic algorithms have been previously used for HEDP applications, with good results \cite{golovkin02,nagayama11,nagayama12}. In particular, there is interest in using multi-objective genetic algorithms to consider several data sets simultaneously (typically 3 or 4). For the 9 data points we consider here, and the even larger sets we aim to use, such multiobjective approaches would be difficult. Our single objective modified $\chi^{2}$ approach is in effect a linear scalarisation of the multiobjective problem and allows much larger datasets to be considered. The trade off is that a single solution is found where multiobjective methods give several candidates; our careful treatment of the error bars on each data point serves to justify our choice of scalarisation.\par
It has been previously noted that the linear model we employ is not justified for ICF targets, since they have been highly tuned to operate at peak performance. The advantages of the analytic expression \eqref{eq:information_likelihood_marginal} are great, and so the authors aim to develop an analytic model that is more suited to ICF data. The linear model does, however, capture the essence of the problem; that complex experiments produce less significant results when compared to existing knowledge.\par
The Bayesian nature of our method allows the consistent analysis of all available data, either by evolving the prior knowledge as more data becomes available or by including all data in a single analysis; the different sets of data do not need to be from the same experiment, or even ones of the same design. These extensions will form a important part of our further work. Finally, the computational methods we have presented are suitable for both experimental design and discovery purposes, and we aim to develop this application.\par
%
%
%
	\refstepcounter{section}									

\begin{thebibliography}{10}

\bibitem{nif}
EL~Moses.
\newblock The national ignition facility and the national ignition campaign.
\newblock {\em IEEE Transactions on Plasma Science}, 38(4):684--689, 2010.

\bibitem{landen12}
O~L Landen, R~Benedetti, D~Bleuel, \emph{Et al.}
\newblock Progress in the indirect-drive national ignition campaign.
\newblock {\em Plasma Physics and Controlled Fusion}, 54(12):124026, 2012.

\bibitem{gaffney12_bayesian_submitted}
JA~Gaffney, D~Clark, V~Sonnad, and SB~Libby.
\newblock Development of a {B}ayesian method for the analysis of {ICF}
  experiments on the {NIF}.
\newblock Submitted to Nuclear Fusion.

\bibitem{hicks12}
D.~G. Hicks, N.~B. Meezan, E.~L. Dewald, \emph{Et al.}
\newblock Implosion dynamics measurements at the national ignition facility.
\newblock {\em Physics of Plasmas}, 19(12):122702, 2012.

\bibitem{castor}
J~Castor.
\newblock {\em Radiation Hydrodynamics}.
\newblock Cambridge University Press, 2004.

\bibitem{haan11}
SW~Haan, JD~Lindl, DA~Callahan, \emph{Et al.}
\newblock Point design targets, specifications, and requirements for the 2010
  ignition campaign on the national ignition facility.
\newblock {\em Physics of Plasmas}, 18:051001, 2011.

\bibitem{atzeni}
S~Atzeni and J~Meyer-ter Vehn.
\newblock {\em The Physics of Inertial Fusion}, volume 125 of {\em Internation
  Series of Monographs on Physics}.
\newblock Oxford Science, 2004.

\bibitem{hanson98}
KM~Hanson and GS~Cunningham.
\newblock Posterio sampling with improved efficiency.
\newblock {\em Proceedings of the SPIE}, 3338:371--382, 1998.

\bibitem{sacks89}
J~Sacks, WJ~Welch, TJ~Mitchell, and HP~Wynn.
\newblock Design and analysis of computer experiments.
\newblock {\em Statistical Science}, 4(4):409--435, 1989.

\bibitem{kennedy01}
Marc~C. Kennedy and Anthony O'Hagan.
\newblock Bayesian calibration of computer models.
\newblock {\em Journal of the Royal Statistical Society: Series B (Statistical
  Methodology)}, 63(3):425--464, 2001.

\bibitem{clark-pres11}
D.~Clark.
\newblock Capsule modeling of {C}on{A}bl {N}101220.
\newblock Presented at NIC Workshop, April 18 2011.

\bibitem{roe07}
Byron~P. Roe.
\newblock Statistical errors in monte carlo estimates of systematic errors.
\newblock {\em Nuclear Instruments and Methods in Physics Research Section A:
  Accelerators, Spectrometers, Detectors and Associated Equipment}, 570(1):159
  -- 164, 2007.

\bibitem{hydra}
M.~M. Marinak, G.~D. Kerbel, N.~A. Gentile, \emph{Et al.}
\newblock Three-dimensional {HYDRA} simulations of national ignition facility
  targets.
\newblock {\em Physics of Plasmas}, 8(5):2275--2280, 2001.

\bibitem{kline10}
J.~L. Kline, K.~Widmann, A.~Warrick, \emph{Et al.} 
\newblock The first measurements of soft x-ray flux from ignition scale
  hohlraums at the national ignition facility using dante.
\newblock {\em Review of Scientific Instruments}, 81(10):10E321, 2010.

\bibitem{hicks10}
D.~G. Hicks, B.~K. Spears, D.~G. Braun, \emph{Et al.}
\newblock Convergent ablator performance measurements.
\newblock {\em Physics of Plasmas}, 17(10):102703, 2010.

\bibitem{golovkin02}
I~Golovkin, R.~Mancini, S~Louis, \emph{Et al.}
\newblock Analysis of {X}-ray spectral data with genetic algorithms.
\newblock {\em Journal of Quanititative Spectroscopy and Radiative Transfer},
  75:625--636, 2002.

\bibitem{nagayama11}
T~Nagayama, R.~Mancini, R~Florido, \emph{Et al.}
\newblock Processing of spectrally resolved x-ray images of intertial
  confinement fusion implosion cores recorded with multimonochromatix x-ray
  imagers.
\newblock {\em Journal of Applied Physics}, 109:093303, 2011.

\bibitem{nagayama12}
T~Nagayama, J.~Bailey, G.~Rochau, \emph{Et al.}
\newblock Investigation of iron opacity experiment plasma gradients with
  synthetic data analyses.
\newblock {\em Review of Scientific Instruments}, 83:10E128, 2012.

\end{thebibliography}
	\bibliographystyle{unsrt}								
%

%
%
	\end{document}